\documentclass[]{spie}  
\usepackage[]{graphicx}
\usepackage{ amssymb }

\title{Performance and on-sky optical characterization of the SPTpol instrument } 

\author{
 E.M.~George\supit{h}, P.~Ade\supit{b}, K.A.~Aird\supit{r}, J.E.~Austermann\supit{c}, J.A.~Beall\supit{d}, D.~Becker\supit{d}, A.~Bender\supit{f}, B.A.~Benson\supit{a,s}, L.E.~Bleem\supit{a,q}, J.~Britton\supit{d}, J.E.~Carlstrom\supit{a,e,q,s,t}, C.L.~Chang\supit{a,e,s}, H.C.~Chiang\supit{a,s}, H-M.~Cho\supit{d}, T.M.~Crawford\supit{a,t}, A.T.~Crites\supit{a,t}, A.~Datesman\supit{g}, T.~de Haan\supit{f}, M.A.~Dobbs\supit{f}, W.~Everett\supit{a}, A.~Ewall-Wice\supit{a,q}, N.W.~Halverson\supit{c,p}, N.~Harrington\supit{h}, J.W.~Henning\supit{c}, G.C.~Hilton\supit{d}, W.L.~Holzapfel\supit{h}, S.~Hoover\supit{a,q}, N.~Huang\supit{a,s}, J.~Hubmayr\supit{d}, K.D.~Irwin\supit{d}, M.~Karfunkle\supit{a,q}, R.~Keisler\supit{a,q,s}, J.~Kennedy\supit{f}, A.T.~Lee\supit{h}, E.~Leitch\supit{a}, D.~Li\supit{d}, M.~Lueker\supit{j}, D.P.~Marrone\supit{o}, J.J.~McMahon\supit{k}, J.~Mehl\supit{a,s}, S.S.~Meyer\supit{a,q,s,t}, J.~Montgomery\supit{a,q}, T.E.~Montroy\supit{m}, J.~Nagy\supit{m}, T.~Natoli\supit{a,q}, J.P.~Nibarger\supit{d}, M.D.~Niemack\supit{d}, V.~Novosad\supit{g}, S.~Padin\supit{a}, C.~Pryke\supit{l}, C.L.~Reichardt\supit{h}, J.E.~Ruhl\supit{m}, B.R.~Saliwanchik\supit{m}, J.T.~Sayre\supit{m}, K.K.~Schaffer\supit{n}, E.~Shirokoff\supit{j}, K.~Story\supit{a,q}, C.~Tucker\supit{b}, K.~Vanderlinde\supit{f}, J.D.~Vieira\supit{j}, G.~Wang\supit{e}, R.~Williamson\supit{a,s}, V.~Yefremenko\supit{e,g}, K.~W.~Yoon\supit{d}, E.~Young\supit{h}
\skiplinehalf
\supit{a} Kavli Institute for Cosmological Physics, Department of Physics, Enrico Fermi Institute, The University of Chicago, Chicago, IL 60637, USA
\skiplinehalf
\supit{b} Cardiff School of Physics and Astronomy, Cardiff University, Cardiff, United Kingdom
\skiplinehalf
\supit{c} Department of Astrophysical and Planetary Sciences, University of Colorado, Boulder, CO 80309, USA
\skiplinehalf
\supit{d} NIST, Boulder, CO 80305, USA
\skiplinehalf
\supit{e} High Energy Physics Division, Argonne National Laboratory, Argonne, IL 60439, USA
\skiplinehalf
\supit{f} McGill University, Montreal, Quebec, Canada
\skiplinehalf
\supit{g} Materials Science Division, Argonne National Laboratory, Argonne, IL 60439, USA
\skiplinehalf
\supit{h} University of California, Berkeley, 151 LeConte Hall Berkeley, CA 94720, USA
\skiplinehalf
\supit{j} California Institute of Technology, Pasadena, CA 91125, USA
\skiplinehalf
\supit{k} University of Michigan, Ann Arbor, MI, USA
\skiplinehalf
\supit{l} University of Minnesota, Minneapolis, MN 55455, USA
\skiplinehalf
\supit{m} Case Western Reserve University, Cleveland, OH 44106, USA
\skiplinehalf
\supit{n} School of the Art Institute of Chicago, Chicago, IL 60603, USA
\skiplinehalf
\supit{o} Steward Observatory, University of Arizona, 933 North Cherry Avenue, Tucson, AZ 85721, USA
\skiplinehalf
\supit{p} Department of Physics, University of Colorado, Boulder, CO 80309, USA
\skiplinehalf
\supit{q} Department of Physics, University of Chicago, 5640 South Ellis Avenue, Chicago, IL 60637, USA
\skiplinehalf
\supit{r} University of Chicago, 5640 South Ellis Avenue, Chicago, IL 60637, USA
\skiplinehalf
\supit{s} Enrico Fermi Institute, University of Chicago, 5640 South Ellis Avenue, Chicago, IL 60637, USA
\skiplinehalf
\supit{t} Department of Astronomy and Astrophysics, University of Chicago, 5640 South Ellis Avenue, Chicago, IL 60637, USA
}

\authorinfo{Further author information: (Send correspondence to Elizabeth M. George.)\\Elizabeth M. George: E-mail: lizinvt@berkeley.edu, Telephone: 1 510 642 5721}

\begin{document} 
  \maketitle 

\begin{abstract}
In January 2012, the 10m South Pole Telescope (SPT) was equipped with a polarization-sensitive camera, SPTpol, in order to measure the polarization anisotropy of the cosmic microwave background (CMB). Measurements of the polarization of the CMB at small angular scales  ($\sim$several arcminutes) can detect the gravitational lensing of the CMB by large scale structure and constrain the sum of the neutrino masses. At large angular scales ($\sim$few degrees) CMB measurements can constrain the energy scale of Inflation. SPTpol is a two-color mm-wave camera that consists of 180 polarimeters at 90 GHz and 588 polarimeters at 150 GHz, with each polarimeter consisting of a dual transition edge sensor (TES) bolometers. The full complement of 150 GHz detectors consists of 7 arrays of 84 ortho-mode transducers (OMTs) that are stripline coupled to two TES detectors per OMT, developed by the TRUCE collaboration and fabricated at NIST. Each 90 GHz pixel consists of two antenna-coupled absorbers coupled to two TES detectors, developed with Argonne National Labs. The 1536 total detectors are read out with digital frequency-domain multiplexing (DfMUX). The SPTpol deployment represents the first on-sky tests of both of these detector technologies, and is one of the first deployed instruments using DfMUX readout technology. We present the details of the design, commissioning, deployment, on-sky optical characterization and detector performance of the complete SPTpol focal plane.
\end{abstract}



\section{Introduction} \label{sec:intro}

The SPTpol instrument is a dual-frequency polarization sensitive camera that was deployed on the South Pole Telescope (SPT), a 10-meter off-axis Gregorian telescope, in January 2012. The instrument is designed for high-angular resolution polarized observations of the cosmic microwave background (CMB). Measurements of CMB polarization at large angular scales yield the tensor to scalar ratio, $r$, which constrains the energy scale of inflation. At small angular scales observations produce maps of weak lensing that constrain the sum of the neutrino masses. SPTpol is among the most sensitive CMB polarization instruments now in operation\cite{bleem12}, and, including \emph{Planck} priors, is expected to have a 1-$\sigma$ uncertainty on the sum of the neutrino masses of $\sigma(\Sigma m_{\nu}) = 0.096$  eV and place a constraint of $\sigma(r) = 0.028$ on the tensor-to-scalar ratio.\cite{austermann12}.

The SPTpol focal plane consists of 1536 total transition edge sensor (TES) bolometers. The full complement of 150~GHz detectors consists of 7 arrays of 84 ortho-mode transducers (OMTs) stripline coupled to two TES detectors each\cite{henning12}. Each of the 180 pixels at 90 GHz consists of two antenna-coupled absorbers coupled to two TES detectors.\cite{sayre12} Both types of detectors are feedhorn coupled to the telescope. The 1536 total detectors are read out with digital frequency-domain multiplexing (DfMUX)\cite{dehaan12}.  The SPTpol deployment represents the first on-sky tests of both of these detector implementations, and some of the first science observations with DfMUX readout technology. 

First light was achieved January 27, 2012, and, after a short commissioning phase which included calibration measurements, CMB observations have been ongoing. These proceedings will present the details of the design, commissioning, deployment, on-sky optical characterization, and detector performance of the first year SPTpol instrument.

\section{Design} \label{sec:design}

\subsection{Telescope and Optics} \label{sec:telescope}

The South Pole Telescope is a 10-m offset classical Gregorian telescope located at the Amundsen-Scott South Pole Station. The optical design consists of two mirrors, a primary and a secondary mirror, and a lens, which creates a telecentric focus. The primary mirror has a 20 $\mu$m rms surface accuracy, and the optical design gives diffraction limited beams at 90 and 150 GHz across a 1.2 degree diameter field of view. This configuration gives low loss, scattering, and instrumental polarization, which is key for observations of the faint polarization signal of the CMB. \cite{carlstrom11}

\begin{figure}
\begin{center}
\begin{tabular}{c}
\includegraphics[width=14cm]{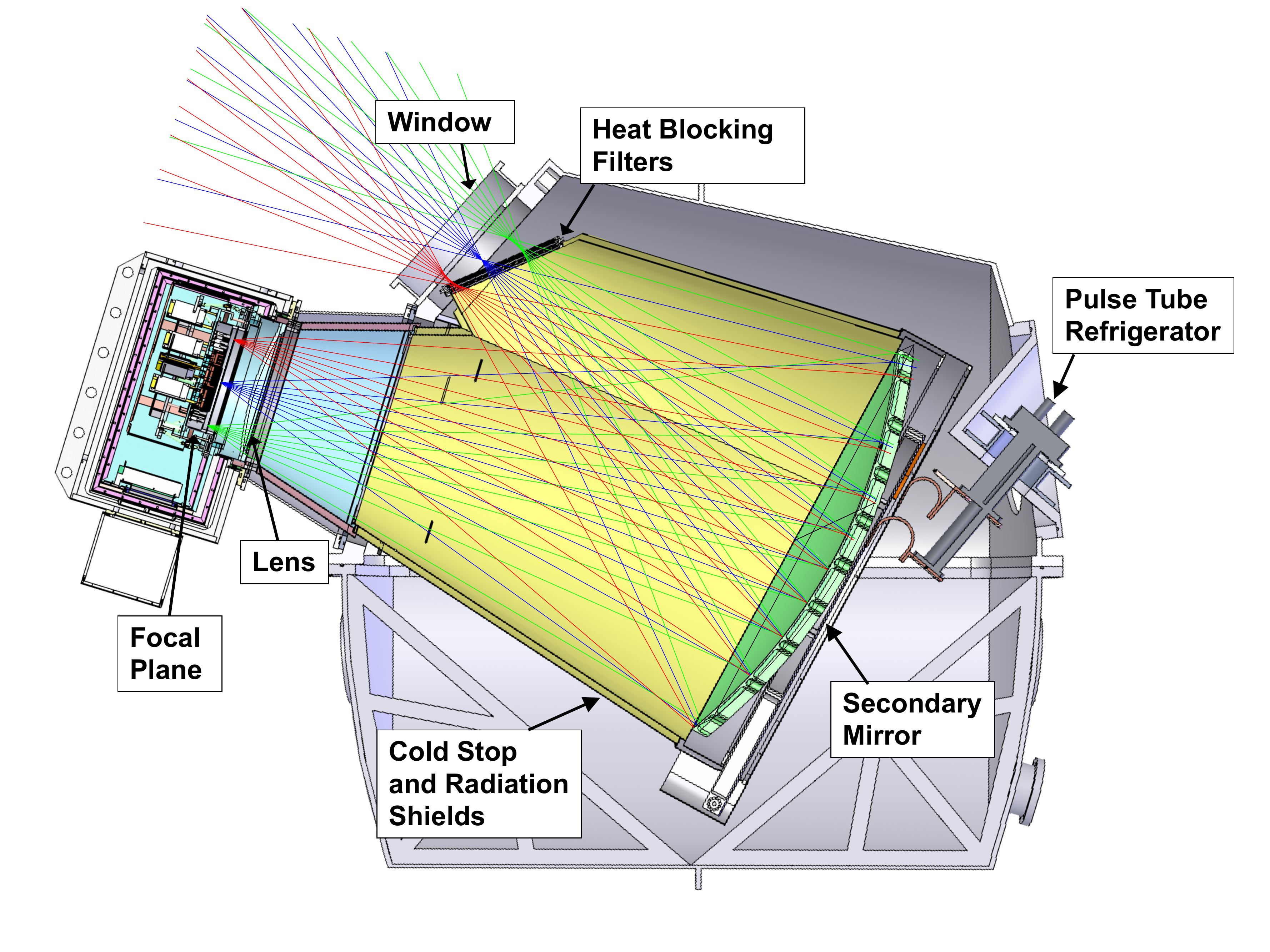}
\end{tabular}
\end{center}
\caption{\label{fig:cryostats} Cutaway of the combined optics cryostat and receiver cryostat with rays entering the vacuum window from the primary mirror (not pictured). The SPTpol instrument replaced the SPT-SZ instrument on the SPT, and the outer shell of the receiver cryostat is an exact duplicate of the SPT-SZ instrument. As such, the same secondary mirror and optics cryostat\cite{carlstrom11} was reused for SPTpol. The thermal filters near prime focus in are tilted and AR coated for SPTpol to prevent polarized reflections from reaching the focal plane.}
\end{figure}

In Figure \ref{fig:cryostats}, we show a cut through the mechanical drawing of the optics and receiver cryostats, with the optical rays over-plotted. The focal plane and secondary mirror are first assembled in separate cryostats, the receiver and optics cryostats, each with its own pulse-tube cooler, which is necessary to cool the secondary mirror and the surrounding cold stop to 10 K. The entrance window to the combined vacuum space is located just before prime focus. Immediately behind the vacuum window, there are a series of infrared (IR) shaders and metal-mesh thermal blocking filters, each tilted with respect to the optical axis to prevent polarized reflections to be reflected back to the focal plane.  These filters are followed by another shader and blocking filter set in front of the lens, and a series of harmonic blocking and band-definiing filters near the focal plane. The thermal filtering is necessary to minimize the heat load on the cold stage, cooled to 280~mK,  and the attached detectors. 

The receiver cryostat has an outer vacuum shell which attaches to the optics cryostat, and two inner shells cooled with a closed-cycle pulse-tube cooler: an intermediate shell cooled to 50~K, and an inner shell cooled to 4~K. On the 4~K stage in the receiver is attached thermal blocking filters and a cold lens that is made out of HDPE and has a broad band AR coating made out of expanded teflon and designed to have $<1\%$ reflection across the 90 and 150~GHz band. A full list of all of the optical elements and the optical load each element contributes is available in Table \ref{tab:loading}. Figure \ref{fig:cryostats} shows the combined optics and receiver cryostats with rays entering from the primary.

\subsection{Receiver Cryogenic Performance} \label{sec:receiver}
 
In addition to the optical elements described in \ref{sec:telescope}, the 4~K stage on the receiver also is the attachment point for the thermal isolation support structure for the focal plane, the SQUIDs (discussed in Section \ref{sec:readout}), and a three-stage $^{4}$He-$^{3}$He-$^{3}$He sorption refrigerator, made by Chase Research, which cools the focal plane to sub-Kelvin temperatures.

\begin{figure}
\begin{center}
\begin{tabular}{c}
\includegraphics[width=7cm]{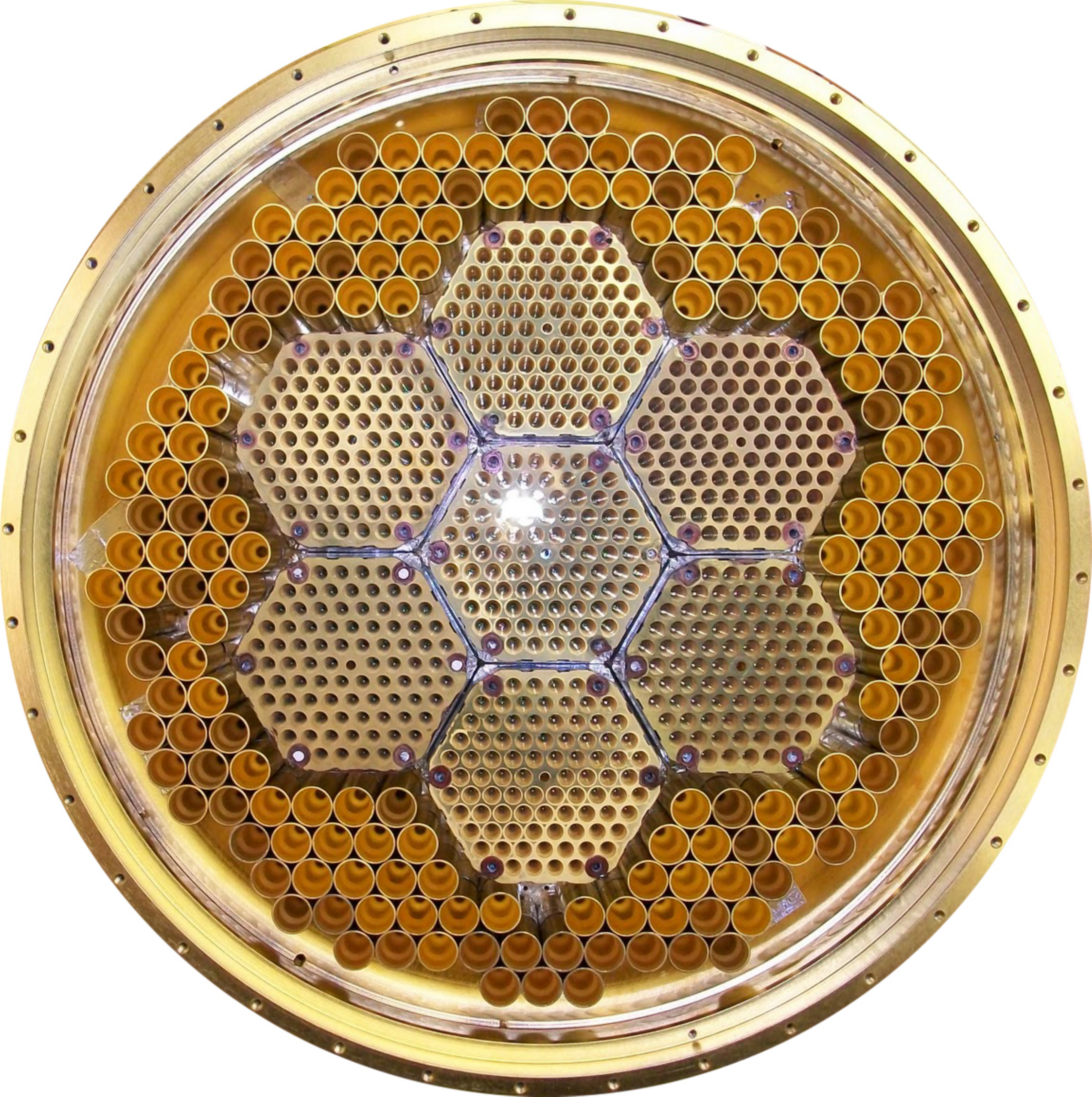}\\
\begin{tabular}{cc}
\includegraphics[width=8cm]{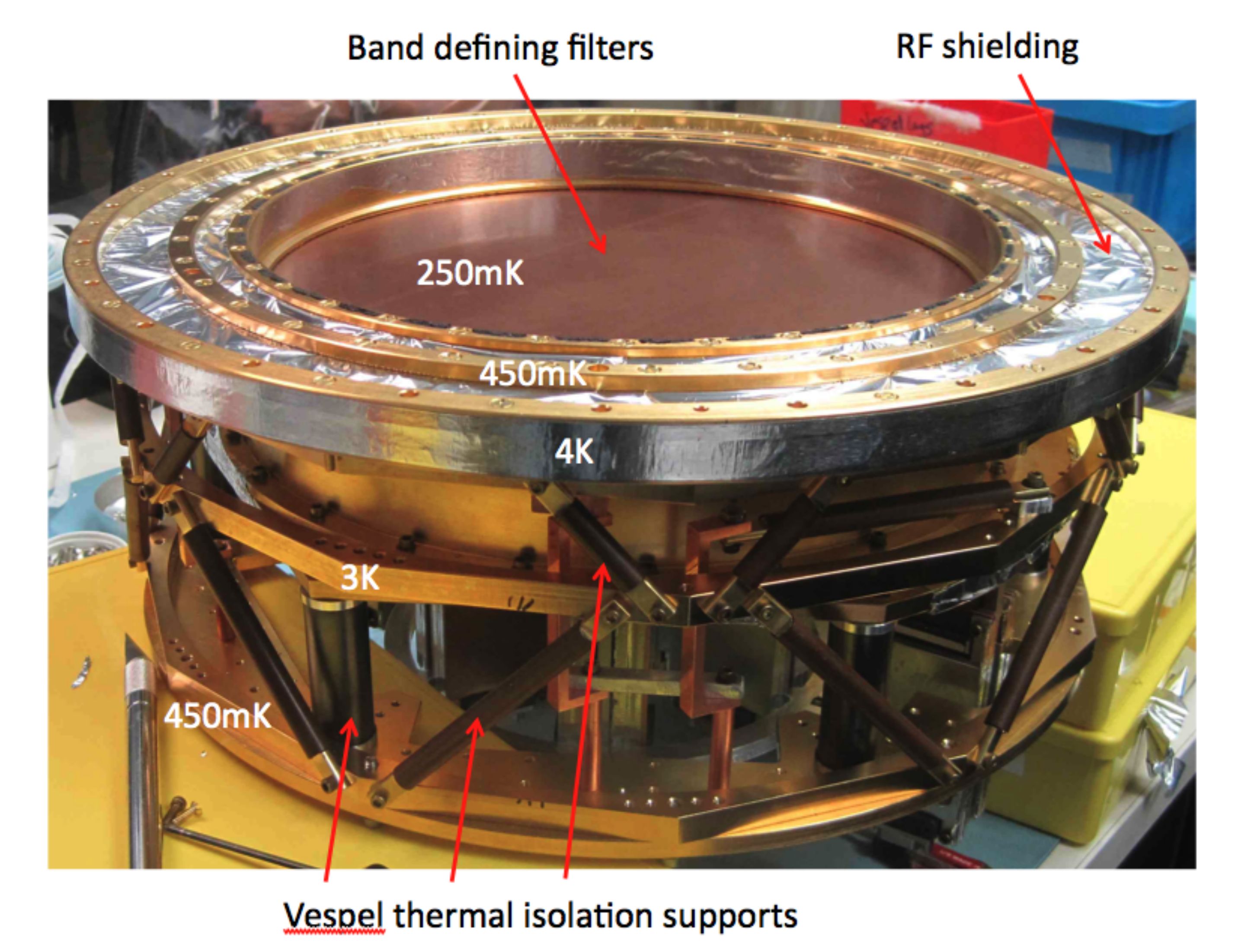} & \includegraphics[width=7.5cm]{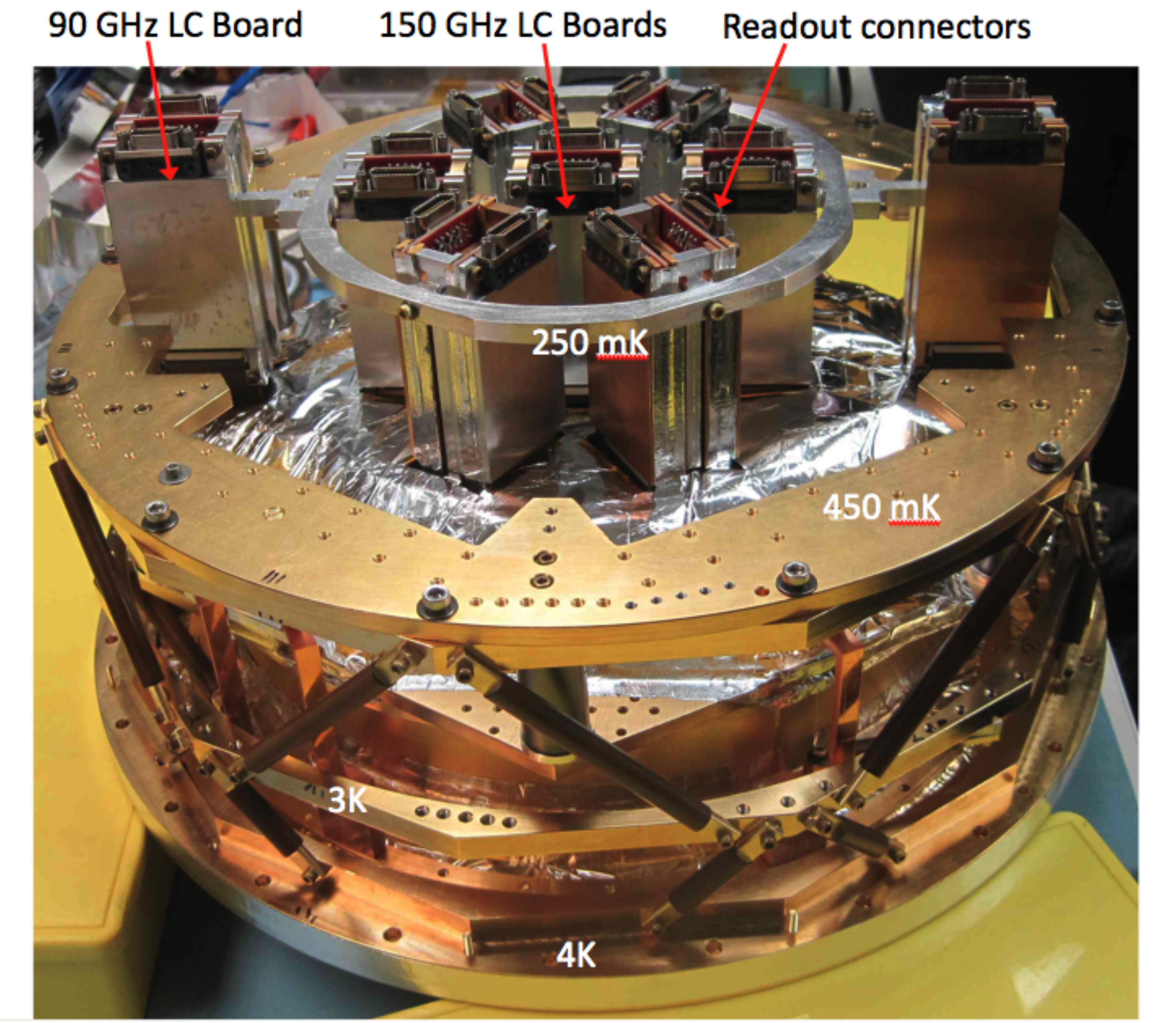} \\
\end{tabular}
\end{tabular}
\end{center}
\caption{\label{fig:focalplane} Pictures of the SPTpol focal plane in various stages of assembly. \emph{Top:} Front side of focal plane with bare horns. 150~GHz arrays are in the center and are surrounded by 90~GHz modules. \emph{Bottom Left:} Front side of focal plane with band-defining filters and RF shielding installed. The outer ring on the top side bolts to the 4~K flange forming the complete RF seal. The Vespel thermal isolation and support legs can be clearly seen in this photo. \emph{Bottom Right:} Back side of focal plane with LC board towers visible. 18 bundles of 8 NbTi twisted pair wires must be routed to the focal plane and heat sunk at 3~K and 450~mK on the way. Heatsink towers not shown.}
\end{figure}

The ultra-cold (UC) stage of the focal plane is isolated from the temperature of the 4~K cryostat shell with thin, hollow Vespel legs that are thermally sunk to intermediate stages of the sorption fridge. The legs are intercepted at 3~K at the Chase heat exchanger (HEX), 450~mK at the Chase inter-cold (IC) stage and finally support the focal plane which is thermally sunk to the 250~mK UC stage. In addition to providing thermal intercepts for the Vespel supports, the HEX and IC stages provide thermal intercepts for the readout wiring (discussed in Section \ref{sec:readout}), and the IC stage supports a thermal blocking filter. The RF shielding for the receiver between the 4~K stage and the cold stage consists of a thin layer of mylar stretched between the two, thermally intercepted at the IC stage. Figure \ref{fig:focalplane} shows the Vespel support structure and thermal intercepts for the focal plane, as well as the RF shielding. 

The combined heat load from Vespel legs, readout wiring, and optical loading puts  $\sim100~\rm{\mu W}$ of power on the IC stage and $\sim6~\rm{\mu W}$ of power on the UC stage. The base temperature of the focal plane during operation is 278 $\pm$ 2~mK. The Chase fridge is cycled approximately every 35 hours with 7 hours spent cycling the fridge and 28 hours for observing. This gives an on sky duty cycle of 80$\%$. 

\subsection{Focal Plane}\label{sec:focalplane}

The focal plane is made up of 1536 TES bolometers mounted to a gold-plated aluminum plate cooled to 278~mK. There are 7$\times$150~GHz arrays, each of which is made up of 84 ortho-mode transducers that are stripline coupled to 2 TES bolometers each for a total of 1176$\times$150~GHz TES bolometers. So that each array individually is sensitive to both Stokes Q and U polarization parameters, the orientation of an OMT is rotated by 45$^{\circ}$ with respect to its neighbors in an alternating pattern.  Each of the 7 arrays is contained in a hexagonal module that includes integrated corrugated silicon horn arrays, the detector array, interface hardware for mounting to the cold stage, and mounts for attaching readout electronics (discussed in Section \ref{sec:readout}). Each module on the focal plane is rotated by 60$^{\circ}$ with respect to its neighbor, giving three independent measurements of Stokes Q and U parameters. A full description of the design of the 150~GHz modules and detector architecture can be found elsewhere in these proceedings.\cite{henning12} The modules are arranged in a hexagonal close-packed arrangement in the center of the focal plane.  

There are 180$\times$90~GHz modules, each of which contains a machined contoured feedhorn and two crossed dipole absorbers, each coupled to a TES, for a total of 360$\times$90~GHz TES bolometers. Each module has 2 alignment pins and 4 electrical bias pins that stick through the mounting plate. The bias pins are soldered to a printed circuit board, which is mounted to the back of the plate and interfaces with the readout electronics. There are 4 independent module angles evenly distributed over the focal plane, each rotated 22.5$^{\circ}$ with respect to each other, giving two independent measurements of Stokes Q and U parameters. The 90~GHz modules and detector architecture are described elsewhere in these proceedings.\cite{sayre12} The 90~GHz modules are arranged in a close-packed ring around the 150~GHz modules. Figure \ref{fig:focalplane} shows the focal plane fully populated with detectors. 

\subsection{Readout}\label{sec:readout}

SPTpol detectors are read out with a digital frequency-domain multiplexed (DfMUX) readout system\cite{dobbs08}, in which the signals from 12 TES bolometers are read out on the same pair of wires. Each TES is biased at a unique frequency between 0.25 and 1.4~MHz at a spacing of roughly 60~kHz. The bias frequencies are set by a 24 $\rm{\mu H}$ inductor and capacitors of various values, placed in series with the TES to form an LCR circuit. The circuit boards that contain the LC components (LC boards) are mounted on the UC stage along with the detector arrays, and are connected to the detectors via superconducting striplines. Further details are provided elsewhere in these proceedings.\cite{henning12}

Each group of 12 TES bolometers is voltage biased by a 30~$\rm{m\Omega}$ shunt resistor in parallel. The voltage bias allows for stable biasing of the detectors in the superconducting transition.\cite{irwin98a} The total power on the TES is constant: $P_{tot} = P_{elec} + P_{opt} = V^2/R_{TES} + P_{opt}$. As the optical power is modulated, the resistance of the TES in the superconducting transition changes, and the resulting change in current is measured using superconducting quantum interference devices (SQUIDs). The SQUIDs are connected to the LC boards through a combination of striplines and NbTi twisted pair wires.  The SQUIDs and bolometers are biased using custom built electronics (McGill DfMUX boards), which also demodulate the bolometer signals. Full details of the DfMUX readout system can be found elsewhere.\cite{dobbs08, smecher12}
 
\section{Detector Properties} \label{sec:detectors}

\subsection{Dark Detector and Operation Parameters} \label{detectorprops}

The overall design for SPTpol detectors is driven by the need to make low noise, polarized measurements of the CMB. Additionally, the details of our readout system and observing conditions impose other requirements. The resistance of the TES bolometers must be matched to our DfMUX readout system, which requires ~$1~\Omega$ TES resistance.\cite{smecher12} This parameter can be tuned by changing the TES geometry. The saturation power of the detectors must be high enough to avoid saturation from optical loading, but remain low enough to minimize the phonon noise contribution.\cite{irwin98a} We tune the saturation power by adjusting the geometry of the thermal link between the TES and bath. The detector time constants must be fast enough to be sensitive to the fine-scale CMB fluctuations as we scan over the sky, but slow enough so that our detectors remain stably biased.\cite{lueker09} We adjust the electrothermal time constants by changing the amount of heat capacity (bling\cite{lueker09}) coupled to the TES. The specific design decisions for the 90 and 150~GHz detectors and measurements of these parameters are covered elsewhere in these proceedings.\cite{sayre12, henning12} 

During CMB observations, the TES detectors are biased in the superconducting transition, where the bias point is defined as a fraction of the normal resistance. The bias point is chosen such that the detectors have high loop gain and remain stable even with small changes in atmospheric loading (see Section \ref{sec:linearity}). The maximum electrical power required to bias the bolometer in the superconducting transition is the detector saturation power minus the optical power (the sources of which are detailed in Table \ref{tab:loading}). The optical time constant is a function of loop gain, and hence where the bolometer is biased in the transition. Measurements of optical time constants are discussed in Section \ref{sec:opticaltau}. Table \ref{tab:detectorprop} lists the relevant detector and operation parameters for the whole focal plane.

\subsection{Uniformity}\label{uniformity}

The uniformity of the detector parameters is different for the 90 and 150~GHz detectors. The 90~GHz detectors were fabricated as single pixels, 25 to a wafer, in many separate fabrication runs. The detectors parameters were generally uniform across a single wafer, but the final focal plane incorporates pixels from many wafers with different parameters. The large standard deviations in the distributions in Table \ref{tab:detectorprop} reflect this.

The 150 GHz detectors were fabricated in 7 arrays of 168 detectors each in two back-to-back fabrication runs. Each array is very uniform in properties, except that two of the 150~GHz detector arrays have a bimodal saturation power distribution. Disregarding the bimodal saturation power distribution, the arrays also have very similar properties to each other. Pre-deployment characterization of 5 of the 7 arrays was carried out, and is described elsewhere in these proceedings.\cite{henning12}

\begin{table}
\begin{center}
\begin{tabular}{|l|c|c|}
\hline
  & \bf{90~GHz} & \bf{150~GHz}\\
 \hline
 Normal Resistance ($R_n$)&$1.0\pm0.1~\Omega$ &$1.2\pm0.2~\Omega$ \\
 \hline
Transition Temp ($T_c$)&$535\pm35~\rm{mK}$ &$478.0\pm28.6~\rm{mK}$ \\
 \hline
 Operation Point&$.74\pm.02~R_n$ &$.78\pm.01~R_n$ \\
 \hline
 Saturation Power at 278mK&$44\pm11~\rm{pW}$ &$22.4\pm5.7~\rm{pW}$  \\
 \hline
 Electrothermal Time Constant&$0.5-1~\rm{ms}$ &$0.5-1~\rm{ms}$ \\
 \hline
 Optical Time Constant&$2.10\pm.78~\rm{ms}$ &$.45\pm.23~\rm{ms}$ \\
 \hline
\end{tabular}
\end{center}
\caption{\label{tab:detectorprop} Detector parameters. The electrothermal time constant listed is for a typical detector deep in the superconducting transition (at high loop gain). The saturation power was measured with the detectors ``dark,'' i.e. with no optical loading. The 150~GHz saturation power distribution is artificially broadened as two of the 150~GHz detector arrays have a bimodal saturation power distributions. Roughly 85$\%$ of the total 150~GHz detectors are in the lower saturation power mode around 20~pW, with only 15$\%$ in the higher saturation power mode around 35~pW.\cite{henning12} }
\end{table} 

\section{Optical Properties}\label{sec:opticalprop}

\subsection{Optical Time Constants}\label{sec:opticaltau}

The signal bandwidth for CMB observations at our current scan speed (0.48 deg/s) lies between 0.1 and 13 Hz, and up to 27~Hz for a 1~deg/s scan speed. The -3~dB point for a 5 ms time constant is $\sim$32 Hz, which is sufficiently fast for the planned SPTpol scan speeds. The readout bandwidth requires our time constants to be slower than approximately 0.25~ms. The optical time constants of the SPTpol detectors are expected to be close to the electro-thermal time constants as there are no large distributed structures that need time to thermalize (as in spiderweb TES bolometers) that would otherwise limit the optical time constants.\cite{wang09} Thus, we require that the optical time constants be $0.25~\rm{ms}<\tau<5\rm~{ms}$.

We measure the optical time constants through a single pole fit to the response to a chopped thermal load at various frequencies.  For the 90 and 150~GHz detectors on a typical observing day, we find the optical time constants to be $2.10 \pm .78~\rm{ms}$ and $.45 \pm .23~\rm{ms}$, which gives a -3~dB point of 75 and 354~Hz respectively. These time constants are more than sufficient for the current scan speed of 0.48~deg/s, and are fast enough to allow a faster scan speed at a future date.

\subsection{Linearity}\label{sec:linearity}

\begin{figure}
\begin{center}
\begin{tabular}{c}
\includegraphics[width=10cm]{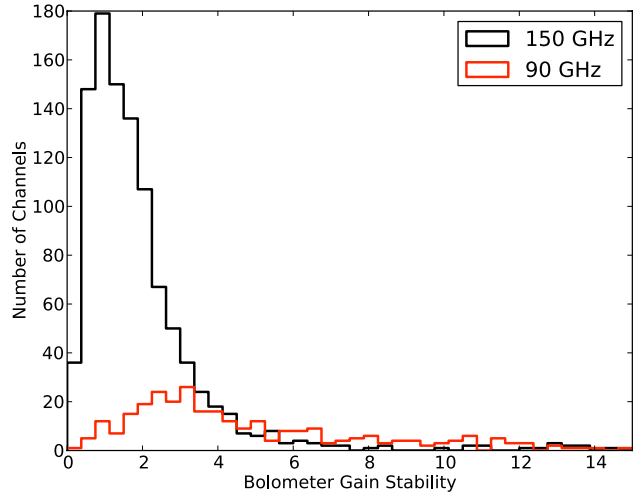}
\end{tabular}
\end{center}
\caption{ \label{fig:response_linearity} Histogram of the gain stability in the response of the detectors at 90 and 150 GHz. Bolometer gain stability is defined as the maximum percent change in response to a chopped calibration source between 45 and 65 degrees.}
\end{figure}

The linearity of the response of the detectors, or gain stability, over a single observing field is important for CMB observations to avoid a systematic mis-calibration across the observation field.  Bolometer gain stability is quantified using the response to a chopped calibrator signal as a function of telescope elevation.  Calibration scans are taken from elevations of 45 to 65 degrees in steps of five degrees.  Each set of scans is fit with a quadratic function and the result normalized to that at an elevation of 55 degrees.  To reduce the effect of atmospheric varations, the median response over several days of observations is used. The resulting bolometer gain stability is defined as the maximum percent change in response over the entire elevation range.  Figure \ref{fig:response_linearity} shows the distribution of stability separated by bolometer frequency. On average, the gain of the 150 and 90~GHz bolometers is stable to $1.5 \pm 1.8\%$  and $3.7 \pm 3.1\%$ over the 20 degree change in elevation, respectively.  This is considered a ``worst case'' metric, as our 2012 observing field is only half this height (10 degrees in elevation). A small number of bolometers have response fluctuations greater than 15$\%$.  In these cases, we attempt to improve the stability by biasing the bolometer further into the superconducting transition.

\subsection{Beams}\label{sec:beams}

\begin{figure}
\begin{center}
\begin{tabular}{c c}
\includegraphics[width=8cm]{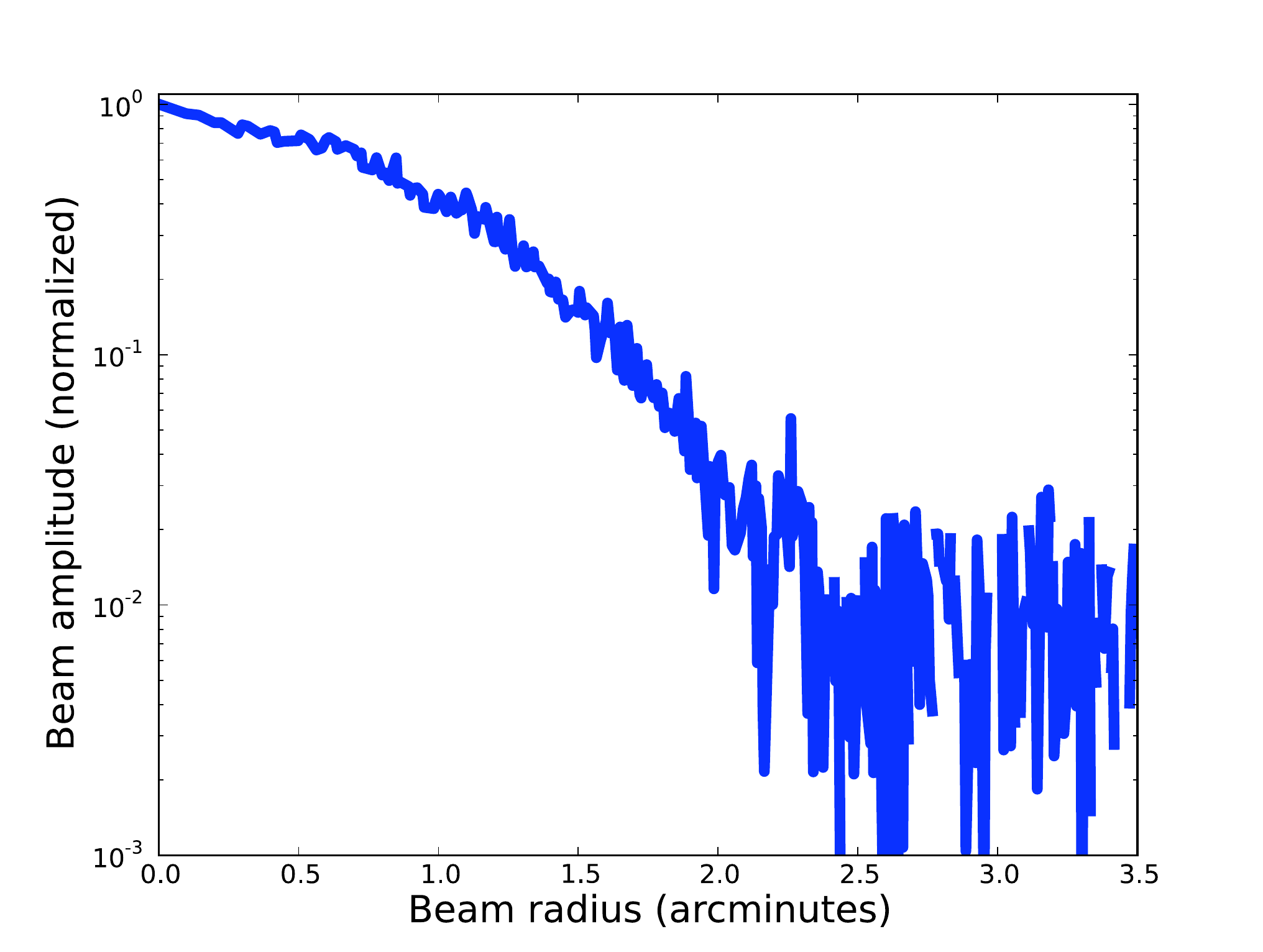}
\includegraphics[width=8cm]{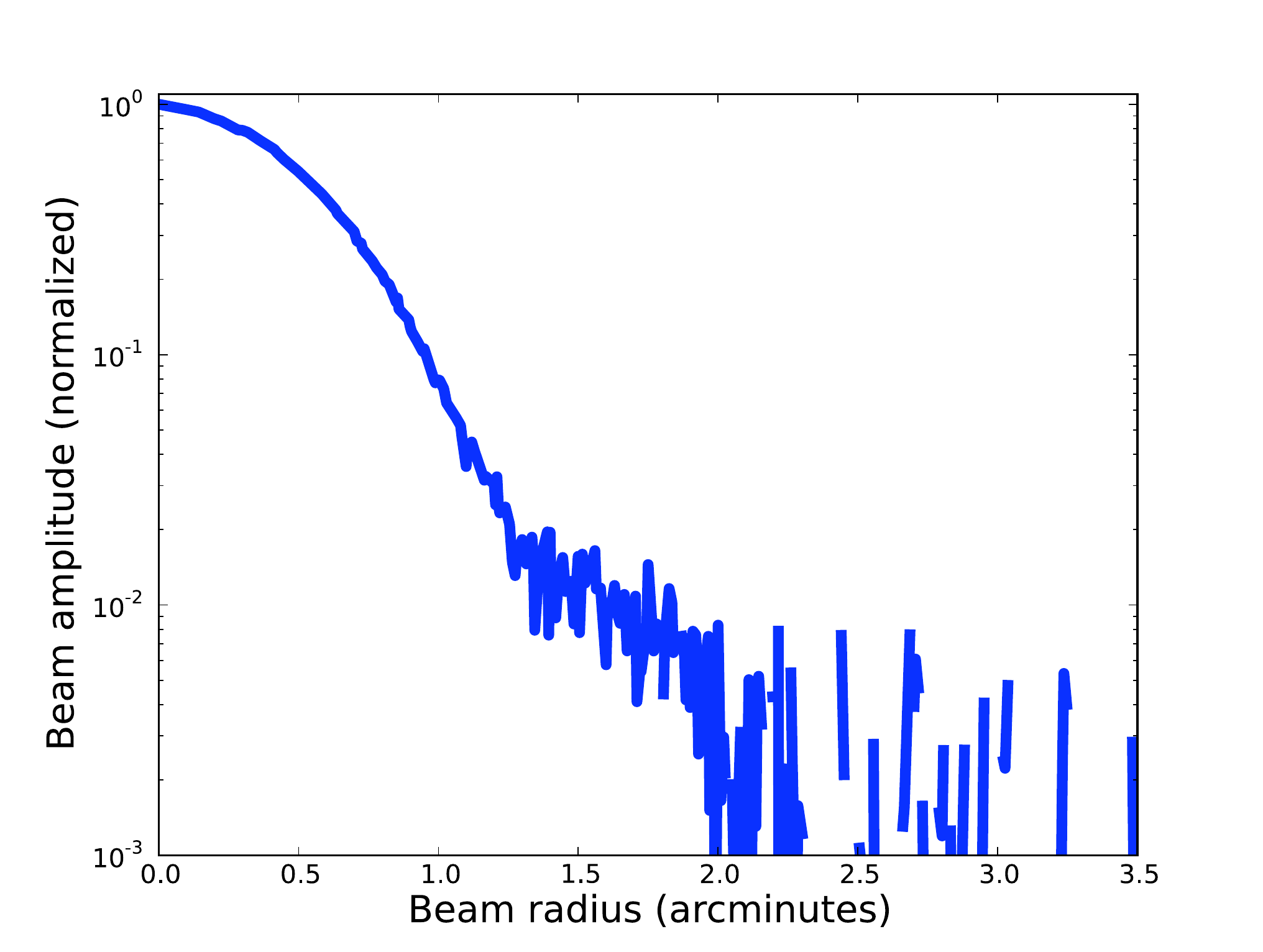}
\end{tabular}
\end{center}
\caption{ \label{fig:beams} Radial profile of 90~GHz (\emph{left}) and 150~GHz (\emph{right}) beams from quasar 0537-441. Beam parameters were fit using an elliptical Gaussian fit to maps of the quasar at each frequency. At 90 and 150 GHz, the FWHM of the beams are 1.83 and 1.06 arcmin, respectively, with a ratio of the minor and major beam axes (a/b) of 0.95 and 0.96, again, respectively.}
\end{figure}

We measure the SPTpol beams out to a large radius using a combination of bright point sources and planets, as was done for SPT-SZ.\cite{lueker10} The next opportunity to do so will be in October 2012 when Mars rises high enough at the South Pole for observation. We have preliminary beams from weekly quasar observations, shown in Figure \ref{fig:beams}.  Beam parameters were fit to maps of the quasar at both frequencies using an elliptical Gaussian. At 90 and 150 GHz, the FWHM of the beams are 1.83 and 1.06 arcmin, respectively, with a ratio of the minor and major beam axes (a/b) of 0.95 and 0.96, again, respectively. Preliminary checks of the beams at different polarization angles show no large differences in shape, with a full analysis pending planet observations in October 2012. The first estimates of pointing offsets using calibration observations of the galactic starforming region RCW38 indicate differential pointing offsets between X and Y polarization in the same pixel to be less than 0.1 arcminute for most pixels.

\subsubsection{Sidelobes}

\begin{figure}
\begin{center}
\begin{tabular}{c}
 \\
\includegraphics[width=11.7cm]{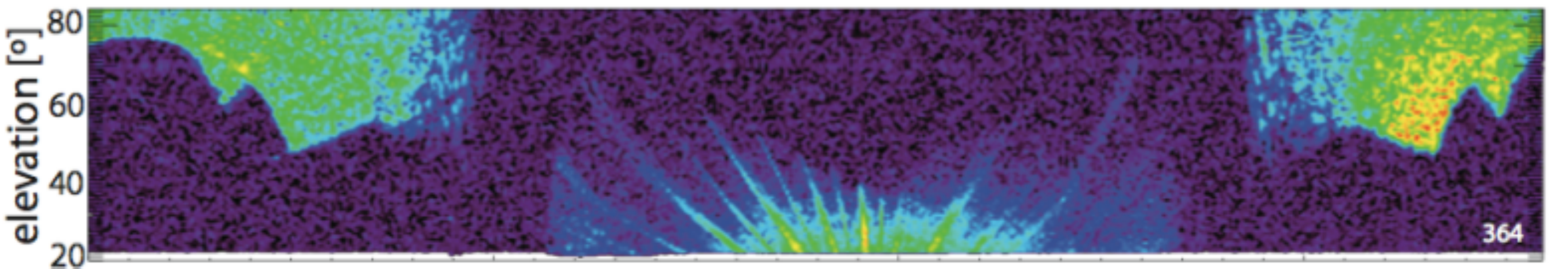}\\
  \includegraphics[width=12.25cm]{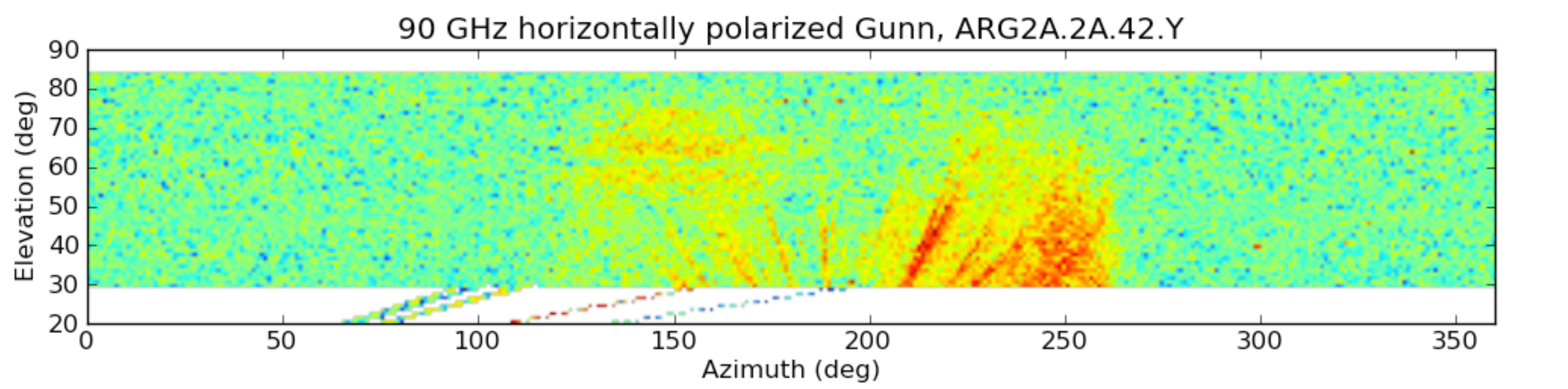}\\
\includegraphics[width=12cm]{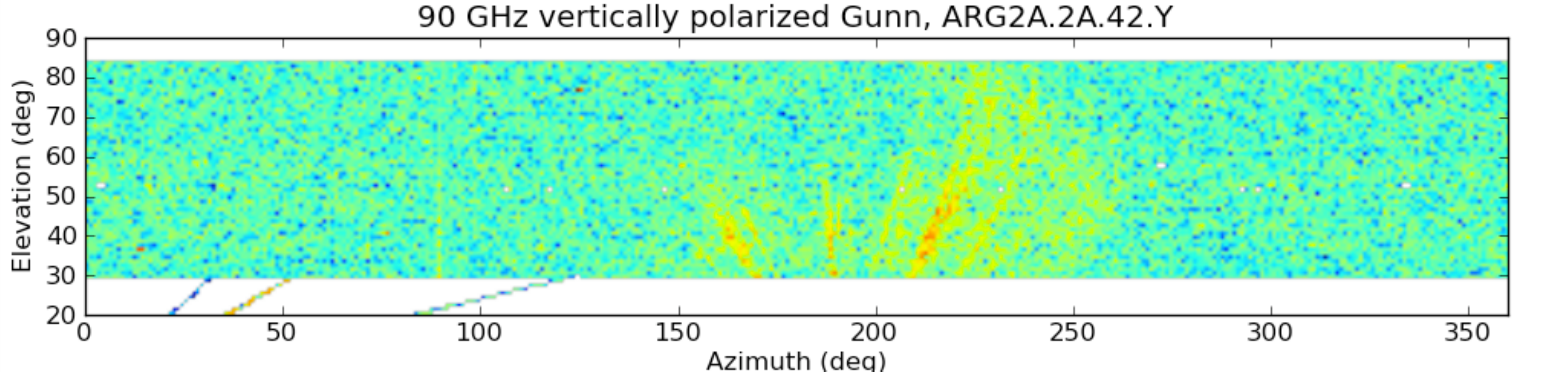}\\
\end{tabular}
\end{center}
\caption{ \label{fig:scattering} Far sidelobe maps made from a single pixel on SPT (top), and a single bolometer of arbitrary polarization on SPTpol (middle and bottom), using a Gunn oscillator source on a nearby building with the first part of the new ground shield installed. SPT and SPTpol maps are on different color scales. The newly installed part of the ground shield blocked all the sidelobe around the primary mirror due to scattering at the cryostat window.  The spiky structure in the middle of the maps (along the main beam) is due to diffraction off of the panel gaps in the primary mirror; it mostly goes to the sky, and the new side shields, to be installed in November 2012, will further shield this lobe from the ground.}
\end{figure}

Along with the installation of the SPTpol camera, we installed the first part of a new co-moving ground shield. The new shield was designed using the measured intensity of the far-sidelobe and scattering beams of SPT.  Highly conservative simulations of the expected ground contamination (i.e., assuming the far-sidelobe and scattering beams are 100$\%$ polarized and that the station buildings and other structures have a contrast of 200 K against the smooth background) indicate that the contamination will be subdominant to the predicted polarization signal for l $\lesssim$ 100, and that using standard observing and analysis mitigation techniques will allow measurement of the signal to much lower multipoles.  Figure \ref{fig:scattering} shows the maps from SPT (top) and SPTpol (middle and bottom) with the first part of the new shield (snout and guard ring) installed. This part of the new shield completely blocks all of the sidelobe around the primary from scattering at the cryostat window. We also plan to fully measure the near sidelobes for the first year configuration in October 2012 when Mars rises high enough to observe. The second part of the ground shield, which further shields the beam from the ground, will be installed in November 2012. 

\subsection{Bandpasses} \label{sec:ftsbands}

The frequency response of the bolometers is defined on the low end by the waveguide cutoff frequency, and on the high end by a metal mesh low-pass filter\cite{ade06}. In addition to band-defining low-pass filters, there is one additional harmonic blocking filter for the 150~GHz pixels and two for the 90~GHz pixels. Bandpasses were measured using a Martin Puplett Fourier transform spectrometer (FTS) located at the entrance to the secondary cryostat, which measures the bandpass of the entire optics chain except the primary mirror. Data were taken for an even sampling of locations on the focal plane, and showed no significant difference in bandpass between different locations. 

\begin{figure}
\begin{center}
\begin{tabular}{c}
\includegraphics[width=13cm]{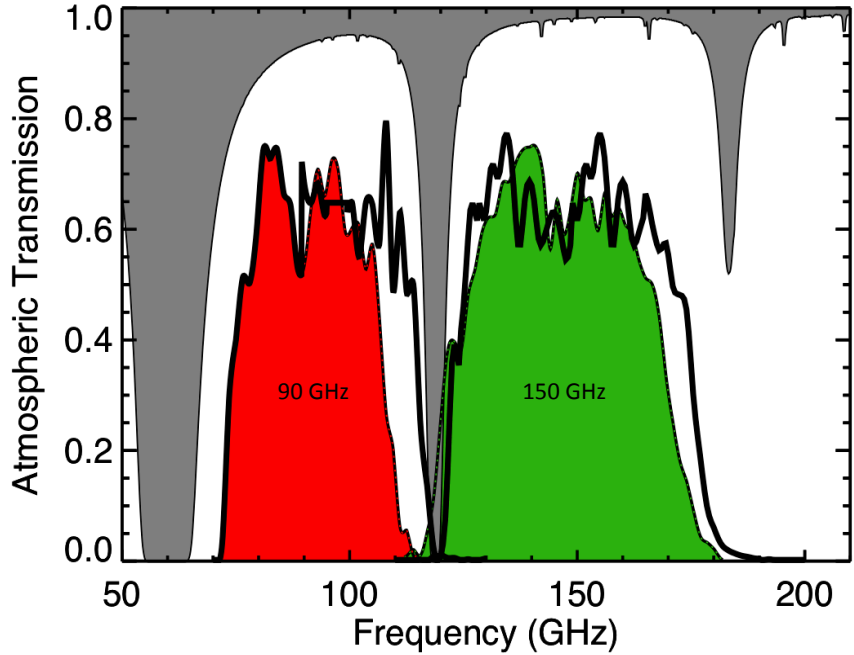}
\end{tabular}
\end{center}
\caption{ \label{fig:ftsbands} Atmospheric transmission model (grey) with measured bandpasses (dashed lines and solid in-fill) and expected bandpasses after the upgrade overlaid (solid lines) for 90 and 150~GHz detectors. Currently the measured 150~GHz bandpass overlaps with the 118~GHz oxygen line, which will be fixed in the upgrade.}
\end{figure}

The measured distribution of the 90 and 150 GHz detectors have respective band centers of $91.4 \pm 1.2$ and $145 \pm 0.8$ GHz and band widths of $30.8 \pm 5.3$ and $43.4 \pm 2.8$ GHz. The full properties of the bandpasses are listed in Table \ref{tab:ftsbands}, and a plot of the bandpasses are shown in Figure \ref{fig:ftsbands}. The current 150~GHz bandpass has a slight overlap with the 118~GHz oxygen line, which causes excess optical loading on the 150~GHz detectors. The 150~GHz bandpass can be optimized by increasing both the high and low end cutoff frequencies, allowing for maximum CMB signal with minimal atmospheric loading. The 90~GHz bandpass can be optimized by increasing the frequency of the high end cutoff to maximize sensitivity to the CMB. We plan to optimize both of the bands in November 2012.

\begin{table}
\begin{center}       
\begin{tabular}{|c|c|c|c|c|c|}
\hline
\bf{Detectors} & \bf{Number} & \bf{Band Center} & \bf{Low Edge} & \bf{High Edge} & \bf{$\Delta \nu$} \\
\hline
90 GHz &62 &$91.2 \pm 1.2$&$73.0 \pm .2 $&$109.7 \pm .2 $&$30.1 \pm 5.2 $\\
\hline
150 GHz &269 &$146.0\pm .8$ &$119.4 \pm .4$ &$172.4 \pm 1.0$ &$43.4 \pm 2.8$\\
\hline
\end{tabular}
\end{center}
\caption{\label{tab:ftsbands} Array averaged bandpasses for 90 and 150~GHz detectors. Data were taken at locations that evenly sampled the focal plane, and showed no significant difference in bandpass between different locations. ``Number'' is the number of pixels measured at each frequency. ``Low edge'' and ``high edge'' are defined as the 25$\%$ transmission point.  After the November 2012 upgrades we expect a 27\% increase in bandwidth in the 90~GHz band, and a slight increase of bandwidth in the 150~GHz band. The band centers for both bands will shift higher in frequency.}
\end{table}

\subsection{Optical Efficiency}\label{sec:opticaleff}

The optical efficiency of the bolometer and horn combination was measured for both the 90 and 150~GHz pixels in the lab through the change in saturation powers of the detectors when a beam-filling cold load was set to different temperatures. The optical efficiency was inferred by $P_{\rm{total}} = \eta P_{\rm{optical}} + P_{\rm{electrical}}$, where $P_{\rm{total}}$ is constant, and $\eta$ is a coupling coefficient where $\eta = \eta_{\rm{horn + bolo}} \times \eta_{\rm{thermal~filters}}$. $P_{\rm{optical}}$ is calculated from the cold load temperature, $P_{\rm{electrical}}$ is measured, and $\eta_{\rm{thermal~filters}}$ can be estimated from the transmission spectra of the filters used during the measurement, leaving $\eta_{\rm{horn+bolo}}$ as the only unknown. For the 90 and 150 pixels, the optical efficiency ($\eta_{\rm{horn+bolo}}$) was measured to be 87$\%$\cite{sayre12} and 90$\%$\cite{henning12} respectively.

The end-to-end optical efficiency of the system can be estimated using the temperature and transmission of each optical element between the bolometer and the sky. Table \ref{tab:loading} gives each optical element, its temperature, emissivity, transmission, and effect on optical efficiency and bolometer optical loading. Correcting for polarization sensitivity and secondary spillover, the predicted band-averaged end-to-end optical efficiency at 90 and 150 GHz is 0.46 and 0.45 respectively. 

The optical efficiency is measured directly on the sky by comparing the brightness of a known astronomical source (RCW38) to the expected brightness. Single pixel maps of RCW38 were fit to templates normalized to 1, converted to Watts, and multiplied by the template area to get $\rm{W \cdot sr}$. To get optical efficiency, this value was compared the expected flux of RCW38 in Janskys, multiplied by the bandwidth and the throughput of the horns. Using this method, the optical efficiency at 90 and 150~GHz was found to be $0.196 \pm 0.046$ and $0.151 \pm 0.017$, respectively. Correcting the polarization sensitivity and secondary spillover, the end-to-end optical efficiency is  $0.45 \pm 0.1$ and $0.48 \pm 0.05$ at 90 and 150~GHz, which is consistent with the expected optical efficiency.

\begin{table}
\begin{center}       
\begin{tabular}{|l|c|c|c|c|c|c|c|}
\hline
\bf{Element} & \bf{$T_e$ [K]} & \bf{$\eta$} & \bf{$L_s$} & \bf{$T_s$ [K]} & \bf{$\eta_{e}$} &\bf{$\eta_{cum}$}&\bf{$P_{opt}$ [pW]} \\
\hline
 Bolometer &.25 &0 (0)&.55 (.55)&.25&.45 (.45)&1.0 (1.0) &1.83e-7 (1.57e-10) \\
\hline
 Cavity &.25 &0 (0)&0 (0) &.25&1.0 (1.0)&.450 (.450) &0 (0)\\
\hline
 Horn &.25 &0 (0)&0 (0) &.25&1.0 (1.0)&.450 (.450)&0 (0)\\
\hline
 Band-def filters&.25 &.050 (.050)&0 (0) &.25&.95 (.95)&.450 (.450) &7.5e-9 (6.44e-12)\\
\hline
 Harmonic blocker&.25 &.050 (.050)&0 (0) &.25&.95 (.95)&.427 (.427) &7.12e-9 (6.12e-12)\\
\hline
 IC blocker&.50 &.050 (.050)&0 (0) &.50&.95 (.95)&.406 (.406)&1.71e-5 (3.86e-7)\\
 \hline
 Lens&6.00 &.020 (.020)&.02 (.02)&10.0&.96 (.96)&.385 (.385)&.076 (.094)\\
\hline 
Lens filter&6.00 &.050 (.050)&0 (0) & 10.0&.95 (.95)&.370 (.370) &.06 (.071)\\
\hline
Secondary &10.0 &.090 (.330)&.02 (.05) &10.0&.88 (.63)&.352 (.352)&.26 (1.09)\\
\hline
 10 K filter&10.0 &.050 (.050)&0 (0) &10.0&.95 (.95)&.313 (.222)&.10 (.09)\\
\hline
 IR shader&100.0 &.010 (.010)&0 (0) &100.0 &.99 (.99)&.297 (.211)&.24 (.25)\\
\hline
 70 K filter&100.0 &..050 (.050)&0 (0) &100.0 &.95 (.95)&.294 (.209)&1.18 (1.23)\\
\hline
 IR shader&100.0 &.010 (.010)&0 (0) &100.0  &.99 (.99)&.279 (.198)&.23 (.233)\\
 \hline
 IR shader&100.0 &.010 (.010)&0 (0) &100.0  &.99 (.99)&.277 (.196)&.22 (.230)\\
\hline
 Window&300.0 &.010 (.010)&0 (0) &300.0 &.99 (.99)&.274 (.194)&.67 (.67)\\
\hline
 Primary&220.0 &.020 (.020)&0 (0) &10.0 &.98 (.98)&.271 (.192)&1.09 (1.21)\\
\hline
 Atmosphere&230.0 &.119 (.074)&0 (0) &230.0 &.88 (.92)&.266 (.189)&5.96 (3.87)\\
\hline
 CMB (peak) &2.73 &1.00 (1.00)&0 (0) &2.73&0 (0)&.234 (.175)&.22 (.13)\\
 CMB (band avg.) &  & & & & &.201 (.143)& \\
\hline
 Total& & & & & & &10.3 (9.17)\\
\hline
\end{tabular}
\end{center}
\caption{\label{tab:loading} Loading table, assuming a 90$\%$ horn + bolo efficiency with a beam filling source for 90 (150)~GHz detectors. $T_e$ is the emission temperature, $\eta$ is the emissivity, $T_s$ is the scattering temperature, $L_s$ is fraction scattered, $\eta_e$ is the efficiency of that element, and $\eta_{cum}$ is the cumulative efficiency. Note that that $\eta_{e}$ is 0.45 for the bolometers. This is because each bolometer is only sensitive to one polarization ($\eta_{e}$ = 0.5), and was measured to be 90$\%$ efficient. The secondary efficiency also includes spillover, so correcting for those two factors gives the actual band averaged end-to-end optical efficiency for 90 and 150~GHz as (0.201/0.88/0.5 = 0.46) and (0.143/0.63/0.5 = 0.45) respectively. }
\end{table}

\subsection{Polarization Properties} \label{sec:polprop}

The polarization properties of the 90 and 150~GHz pixels and horn arrays were measured in the lab before deployment. A test of two stacked 90~GHz detectors shows a cross-polarization response of $< 1.6 \%$, and a relative misalignment between the antennas of $<1^{\circ}$. \cite{sayre12, chang12} The 150~GHz detectors show a cross-polarization response of $0.3\%$. At 150~GHz, the two polarizations are not prone to alignment rotations as the alignment is defined lithographically on a common plane.\cite{henning12}

To measure the polarization angles of the complete array as installed in the telescope, a polarization calibration source placed 3~km from the telescope is employed. The polarization calibration source consists of a chopped blackbody at $\sim1000^{\circ}$C with a fixed polarized grid and a rotating polarized grid to define and vary the polarization. The source is placed in a hole cut in the center of a 24' x 24' reflector angled at 60$^{\circ}$ to the horizon, such that when the telescope is scanned over the reflector, most of the beam is reflected to the sky. For each pixel position, the rotating grid steps through a set of discrete angles spaced at 15$^{\circ}$ to measure the polarization angle of the two bolometers. A preliminary analysis suggests that the polarization angles are measured with a precision of $\sim1^{\circ}$
using this technique. Analysis of the polarization calibration is ongoing, and further details about the polarization calibration can be found elsewhere in these proceedings.\cite{natoli12, austermann12}

\subsection{Array Optical Yield}\label{yield}

During every fridge cycle the detectors are biased to their operation point in the superconducting transition. We run a series of diagnostic observations to check the detector performance, and use the results of these tests to eliminate poorly behaving detectors from our CMB observations. One metric we use is the response of the detectors to a short dip in elevation, and another is measuring the response of the detectors to a chopped thermal load at 6~Hz. If the signal to noise during these diagnostic tests is greater than 10, we consider the bolometer to be operational. On a typical observing day, roughly 80$\%$ of the detectors at both 90 and 150~GHz pass these diagnostic tests. 

Currently we have $\sim80\%$ array yield, but we can do better. We are continuously improving array yield during day-to-day observations by optimizing the operation parameters of under-performing detectors. In the first year configuration we also have a few readout components (such as SQUIDs) that are non-functional or operating at an increased noise level. During the first year we are identifying these components, which will be replaced in the November 2012 upgrade. We can expect a several percent increase in yield from these changes, and a commensurate increase in mapping speed.

\section{Noise}\label{sec:noise}

Given the bolometer optical efficiency, detector saturation powers, bandpasses, expected optical load, and readout noise, the individual bolometer noise equivalent temperatures (NETs) can be predicted. At 90 and 150~GHz, the NETs are predicted to be 501~$\rm{\mu K\sqrt{s}}$ and 517~$\rm{\mu K\sqrt{s}}$ respectively. After the bandpass upgrade in November 2012, we expect the NETs at 90 and 150~GHz to drop to 442~$\rm{\mu K\sqrt{s}}$ and~469 $\rm{\mu K\sqrt{s}}$ respectively, which corresponds to an increase of mapping speed of 25$\%$ and 17$\%$. Table \ref{tab:prednoise} breaks down the contributions the the expected detector NETs.

We get a preliminary measure of the first year noise performance by comparing the signal to noise of the detector response to an astronomical source with a known brightness temperature. We scanned the array over a celestial calibration source, the starforming region RCW38, and fit the data to a template to determine the brightness. We measured the noise in the 9-11~Hz bin during a 5 minute noise stare directly before the RCW38 observation. Using this method, distributions of detector NETs have a mode around 580~$\mu K \sqrt{s}$ for both 90 and 150~GHz, which is consistent with our expected NETs for the first year given 10-20$\%$ calibration uncertainties at this time. 

Given the mode of the noise distribution (580~$\rm{\mu K \sqrt{s}}$) and the typical number of operational bolometers on any day (80$\%$), we have a preliminary estimate of focal plane NET for the first year of 34~$\rm{\mu K \sqrt{s}}$ at 90~GHz and~19 $\rm{\mu K \sqrt{s}}$ at 150 GHz. We expect to eventually achieve an absolute calibration uncertainty of $\lesssim 2\%$ using the measured CMB power spectrum from either WMAP or Planck, which will improve the accuracy of our sensitivity estimates. Typical noise spectra for individual 90 and 150~GHz detectors can be found elsewhere in these proceedings.\cite{sayre12, henning12}

\begin{table}[h]
\begin{center}       
\begin{tabular}{|l|r|r|r|r|r|r|r|r|r|r|}
\hline
  & & & & & &NEP &NEP &NEP &NEP &NET$_{bolo}$ \\  
Band &Center &Width &$P_{opt}$ &$P_{atm}$ &$P_{sat}$ &Photon &Phonon &Current &Total &Total\\
(GHz) & (GHz) & (GHz) & (pW)  & (pW) &(pW) &($\frac{aW}{\sqrt{Hz}}$) &($\frac{aW}{\sqrt{Hz}}$) &($\frac{aW}{\sqrt{Hz}}$)&($\frac{aW}{\sqrt{Hz}}$) &($uK \sqrt{s}$)\\
\hline
\bf{Year 1}\\
\hline
90 &91.2 & 30.1& 10.2 & 5.97 & 40 & 68.6 & 48.5 & 45.3 & 95.5 & 501\\
\hline
150 & 146.0 & 43.4 & 9.17 & 3.87 & 22 & 60.5 & 36.0 & 29.7 & 76.4 & 517\\
\hline
\bf{Year 2}\\
\hline
90 & 94.1 & 37.8 & 13.9 & 8.41 & 40 & 82.7 & 48.5 & 42.3 & 104.9 & 442\\
\hline
150 & 148.5 & 45.5 & 8.22 & 2.79 & 22 & 55.6 & 36.0 & 30.8 & 73.1 & 469\\
\hline
\end{tabular}
\end{center}
\caption{\label{tab:prednoise} Predicted bolometer NETs based on bolometer optical efficiency (assumed to be 90$\%$), detector saturation powers, bandpasses, and expected optical load for 90 and 150~GHz for the current configuration, as well as predictions for noise levels after the November 2012 upgrade. This assumes the photon bunching term is 1.0 and the thermal conduction has a power of n=3 (consistent with silicon nitride) for the phonon noise. If we assume bunching is 0.5, this decreases the predicted total NEP and NET by factors of 1.141 and 1.077 at 90 and 150~GHz, respectively.}
\end{table}

\section{Conclusion}\label{sec:conclusion}

In January 2012, the SPTpol instrument was deployed on the South Pole Telescope. This dual-frequency 1536 TES focal plane is currently one of the most sensitive polarized CMB experiments. Preliminary characterization of the instrument in its first year configuration reveals beams, polarization properties, linearity, optical efficiency, and noise properties consistent with our expectations. In a typical observing day 80$\%$ of detectors show good optical response with the mode of the NET distribution close to the predicted value. Throughout the first year, we will continue to optimize detector performance during day to day operation. A planned upgrade to optimize the bandpasses and replace some readout components in November 2012 will increase detector mapping speed by at least 25$\%$ and 17$\%$ at 90 and 150~GHz.

\acknowledgments     
 
Work at the University of Colorado is supported by the NSF through grant AST-0705302.  Work at NIST is supported by the NIST Innovations in Measurement Science program.  The McGill authors acknowledge funding from the Natural Sciences and Engineering Research Council, Canadian Institute for Advanced Research, and Canada Research Chairs program. MD acknowledges support from an Alfred P. Sloan Research Fellowship.  Work at the University of Chicago is supported by grants from the NSF (awards ANT-0638937 and PHY-0114422), the Kavli Foundation, and the Gordon and Betty Moore Foundation. Work at Argonne National Lab is supported by UChicago Argonne, LLC, Operator of Argonne National Laboratory (``Argonne''). Argonne, a U.S. Department of Energy Office of Science Laboratory, is operated under Contract No. DE-AC02-06CH11357. We acknowledge support from the Argonne Center for Nanoscale Materials.


\bibliography{spiegeorge}   
\bibliographystyle{spiebib}   

\end{document}